\documentclass[a4paper]{jpconf}
\usepackage{graphicx}
\usepackage[spanish,english]{babel} 
\usepackage[utf8]{inputenc} 
\usepackage[T1]{fontenc} 
\usepackage{lineno}
\begin{document}
\title[Simulating the nanometric track-structure of carbon ions in liquid water]{Simulating the nanometric track-structure of carbon ion beams in liquid water at energies relevant for hadrontherapy
}

\author{Pablo de Vera$^1$, Stefano Simonucci$^2$, Paolo E. Trevisanutto$^{3,4}$, Isabel Abril$^5$, Maurizio Dapor$^{3,4}$, Simone Taioli$^{3,4,6}$ and Rafael Garcia-Molina$^1$
}

\address{$^1$Departamento de Física - Centro de Investigación en Óptica y Nanofísica (CIOyN), Universidad de Murcia, 30100 Murcia, Spain}
\address{$^2$School of Science and Technology, University of Camerino, 62032 Camerino, Italy, and INFN, Sezione di Perugia, 06123 Perugia, Italy}
\address{$^3$European Centre for Theoretical Studies in Nuclear Physics and Related Areas (ECT*-FBK), 
38123 Trento, Italy}
\address{$^4$Trento Institute for Fundamental Physics and Applications, 
38123 Trento, Italy}
\address{$^5$Departament de Física Aplicada, Universitat d’Alacant, 03080 Alacant, Spain}
\address{$^6$Peter the Great St. Petersburg Polytechnic University, Russia}
\ead{pablo.vera@um.es}


\begin{abstract}
The nanometric track-structure of energetic ion beams in biological media determines the direct physical damage to living cells, which is one of the main responsibles of their killing or inactivation during radiotherapy treatments or under cosmic radiation bombardment. In the present work, detailed track-structure Monte Carlo simulations, performed with the code SEED (Secondary Electron Energy Deposition), are presented for carbon ions in a wide energy range in liquid water. Liquid water is the main constituent of biological tissues, and carbon ions are one of the most promising projectiles currently available for ion beam cancer therapy. The simulations are based on accurate cross sections for the different elastic and inelastic events determining the interaction of charged particles with condensed-phase materials. The latter are derived from the \textit{ab initio} calculation of the electronic excitation spectrum of liquid water by means of time-dependent density functional theory (TDDFT), which is then used within the dielectric formalism to obtain inelastic electronic cross sections for both carbon ions and secondary electrons. Both the ionisation cross sections of water by carbon ions and the excitation and ionisation cross sections for electron impact are obtained in very good agreement with known experimental data. The elastic scattering cross sections for electrons in condensed-phase water are also obtained from \textit{ab initio} calculations by solving the Dirac-Hartree-Fock equation. The detailed simulations fed with reliable cross sections allow to assess the contribution of different physical mechanisms (electronic excitation, ionisation and dissociative electron attachment --DEA--) to the carbon ion-induced direct biodamage.
\end{abstract}

\section{Introduction}

Among the many applications of ion beams interaction with materials, the irradiation of living tissues with energetic carbon ions is attracting much attention for its benefits for treating deep-seated tumours in the technique known as hadrontherapy \cite{Schardt2010,Ebner2016}. Due to the characteristic track-structure of carbon ions (as compared to that of photon beams used in conventional radiotherapy) \cite{Nikjoo2012}, they produce large clusters of inelastic events within a few turns of the sensitive DNA molecules, thus presenting a large relative biological effectiveness or cell-killing ability.

The development of accurate biophysical models requires the detailed simulation of carbon ions track-structure in liquid water (the main constituent of biological tissue) in a wide energy range, from the MeV/u energies of the primary therapeutic beams down to the keV/u energies at which the stopping power and biological effects are largest. For this purpose, probability distributions (cross sections) of electronic excitations by both carbon ions and their secondary electrons (especially for energies below 50--100 eV, as these electrons are produced in large numbers by ion-impact ionisation) need to be accurately known, in order to be used within Monte Carlo simulation codes for the transport of radiation.

Reliable inelastic electronic cross sections can be obtained by means of the dielectric formalism \cite{Lindhard1954,Ritchie1959} for both ions \cite{deVera2013PRL,deVera2015} and electrons \cite{deVera2019JChemPhysC,deVera2021} in liquid water if its electronic excitation spectrum is accurately known. The latter was determined experimentally for relatively wide energy and momentum transfer ranges \cite{Watanabe1997,Hayashi2000}, although theoretical models are useful to confirm the experiments, interpret them, as well as to fill the gap where experiments are missing. Phenomenological approaches such as the Mermin Energy Loss Function -- Generalised Oscillator Strengths (MELF-GOS) method have been shown to provide reliable results \cite{HerediaAvalos2005PRA,GarciaMolina2011}. However, more recently, time-dependent density functional theory (TDDFT) methods have demonstrated their ability to achieve higher accuracies from purely \textit{ab initio} calculations \cite{Taioli2021JPCL}. Still, TDDFT is too expensive to be applied for arbitrarily large energies or momenta, making analytical approaches such as MELF-GOS necessary to complement the \textit{ab initio} calculations \cite{Pedrielli2021}. 

In this contribution, we show how the TDDFT electronic excitation spectrum can excellently
reproduce the available experimental data up to 100 eV \cite{Watanabe1997,Hayashi2000} and, from it and a high energy Aand high momenta extension by means of the MELF-GOS method, we calculate differential and total cross sections for electronic excitation and ionisation 
in very good agreement with the most recent experimental information.
These cross sections are then used within the Monte Carlo code SEED (Secondary Electron Energy Deposition) \cite{Taioli2021JPCL,Dapor2017PRB,DaporBook}, in order to assess the clustering of direct damaging events produced by carbon ions, in a wide energy range, in nanocylinders of volume equivalent to two DNA convolutions located at different distances from the ions' paths. From the simulations, the role of different physical mechanisms to the clustered nanometric biodamage can be assessed. The Monte Carlo simulations can also be checked against experimental nanodosimetric distributions for ionisation clusters, this interaction being the one which can be most easily measured in the laboratory \cite{Conte2018}.





\section{Excitation spectrum of liquid water}

\begin{figure}[t]
	\includegraphics[width=\columnwidth]{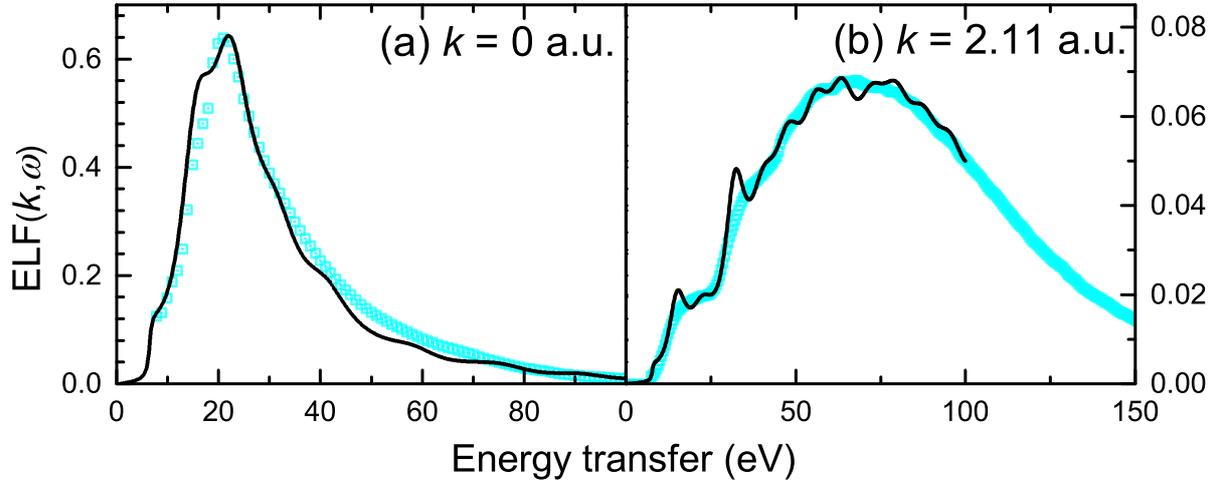}
	\caption[]{Energy loss function of liquid water as a function of the energy transfer $E = \hbar \omega$ for momentum transfers (a) $\hbar k=0$ a.u. and (b) $\hbar k=2.11$ a.u.. Symbols correspond to experimental determinations for liquid water \cite{Watanabe1997,Hayashi2000}, while lines depict present TDDFT calculations.}
	\label{fig:ELF}
\end{figure}

The energy loss function (ELF) of liquid water, Im$[-1/\epsilon(k,\omega)]$, gives the excitation spectrum over the energy $\hbar \omega$ and momentum $\hbar k$ transfers surface, $\epsilon(k,\omega)$ being the complex dielectric function. The ELF is obtained here by TDDFT, using the Lanczos chains algorithm (LCA) implemented in the turboEELS code \cite{Turbo-eels} and the adiabatic PBE (APBE) kernel~\cite{Olsen2019}, over the energy range $0 \le E\le 100 \mbox{ eV}$ and for momentum transfers $0 \le \hbar k \le 2.5 \mbox{ a.u.}$, with a resolution of 0.25 a.u. These calculations were done for a particular configuration of a periodic box formed by 32 water molecules and with liquid water density, electronically relaxed via DFT calculations in Quantum Espresso \cite{Giannozzi2009} using the PBE-GGA functional \cite{Olsen2019} and Troullier-Martins (TM) norm-conserving pseudopotentials. Such a box was carved out from a supercell of several thousands molecules optimised by means of molecular dynamics simulations using the empirical TIP3P force-field \cite{MacKerell1998} implemented in LAMMPS \cite{LAMMPS2020}. Figure \ref{fig:ELF} depicts by solid lines the TDDFT results for the ELF of liquid water for two exemplary momentum transfers of (a) 0 a.u. and (b) 2.11 a.u., together with the available experimental data (symbols) \cite{Watanabe1997,Hayashi2000}. The \textit{ab initio} calculations yield excellent results, reproducing very well the shape and intensity of the different peaks corresponding to the excitations of the valence electrons of liquid water.

\section{Ionisation cross sections for carbon ions in liquid water}

\begin{figure}[h!]
	\includegraphics[width=0.51\columnwidth]{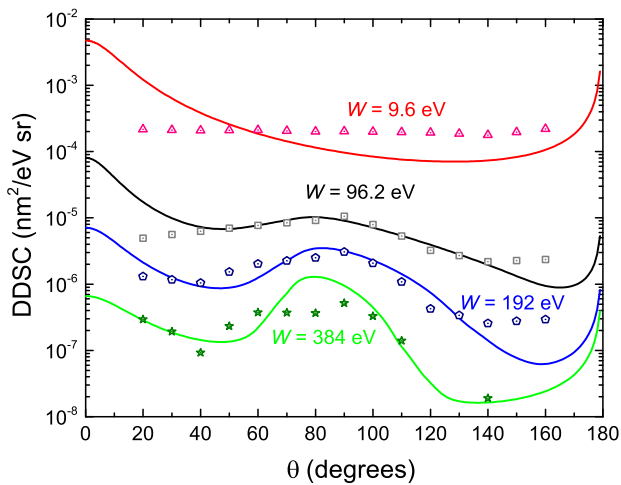}\hspace{2pc}%
	\begin{minipage}[b]{14pc}\caption{\label{fig:DDCS_C}Angular distributions of secondary electrons ejected from water at different kinetic energies $W$ for the impact of 6 MeV C$^{6+}$ ions. Symbols correspond to experimental data for the water molecule \cite{DalCappello2009}, while lines are calculations for liquid water based on the ELF obtained from TDDFT.}
	\end{minipage}
\end{figure}

The space and energy characteristics of the ion track-structure are determined by the angular and energy distributions of the secondary electrons which are put in motion by the primary beam.
Through the dielectric formalism, the ELF over the entire ($k$, $\omega$) plane allows the calculation of the angular and energy spectra of the electrons produced by carbon ions in liquid water, following the model presented in Refs. \cite{deVera2013PRL,deVera2015}. In short, the excitation and ionisation of the valence shell electrons are distinguished by defining a mean binding energy for these electrons, which for liquid water has a value of 13.71 eV \cite{deVera2021}. The $\omega$-dependence of the ELF gives the energy spectrum of secondary electrons \cite{deVera2013PRL}, while the $k$-dependence, through a physically motivated relation, provides their angular dependence \cite{deVera2015}. The calculated angular distributions of secondary electrons produced by 6 MeV/u C$^{6+}$ ions in liquid water are shown in Fig. \ref{fig:DDCS_C}, for several kinetic energies of the secondary electrons $W$ (i.e., the doubly differential cross sections, DDCS), together with the experimental measurements available for water molecules \cite{DalCappello2009}. 

Despite the expected phase effects, in general the calculations reproduce very well the experimental points recorded in water vapour in a wide range of energies and angles, confirming the reliability of the model. For the lower energies, the calculations predict a forward peaked distribution in opposition to the experiments, a difference which may be due to phase effects. The figure exemplifies how low energy electrons are produced in much larger quantity, while higher energy electrons tend to be ejected at the angle defined by the binary encounter peak.


\section{Cross sections for secondary electrons in liquid water}

The dielectric formalism also allows the calculation of electron inelastic scattering cross sections for liquid water. For that purpose, the model needs to be amended for indistinguishability and exchange between the primary and excited electrons, both for electronic excitation and ionisation. Furthermore, electronic structure effects and low-energy corrections need to be considered in order to describe the energy-loss of secondary electrons \cite{deVera2021}.

\begin{figure}[t]
	\includegraphics[width=\columnwidth]{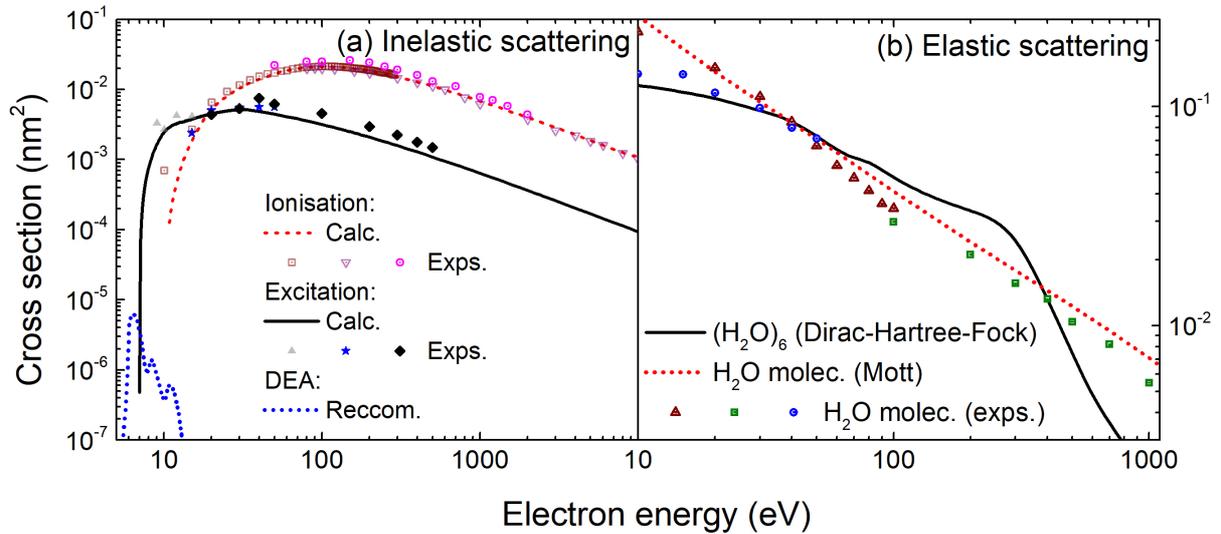}
	\caption[]{Total cross sections for electron impact in water. (a) Inelastic cross sections: Solid and dashed lines correspond to present calculations for electronic excitation and ionisation using the ELF obtained from TDDFT, the dotted line corresponds to the recommended DEA cross section \cite{Itikawa2005}, open symbols depict experimental data for ionisation (down triangles \cite{Schutten1966}, circles \cite{Bolorizadeh1986}, squares \cite{Bull2014}) and full symbols for excitation (diamonds \cite{Thorn2007JCP}, triangles \cite{Ralphs2013}, stars \cite{Matsui2016}). (b) Elastic cross section: the solid lines corresponds to the Dirac-Hartree-Fock calculation for a cluster of six water molecules, the dotted line to the Mott calculation for the water molecule, while symbols correspond to experimental data for the water molecule (triangles \cite{Itikawa2005}, squares \cite{Katase1986}, circles \cite{Cho2004a}).}
	\label{fig:XSe}
\end{figure}

Figure \ref{fig:XSe}(a) represents by solid (dashed) lines the electronic excitation (ionisation) cross sections calculated for liquid water. Symbols depict a compilation of experimental data which, unfortunately, are only available for water vapour. For ionisation, the calculations reproduce very well the experimental points \cite{Schutten1966,Bolorizadeh1986,Bull2014} in the entire energy range, particularly the measurements from Refs. \cite{Schutten1966,Bull2014}. Those from Ref. \cite{Bolorizadeh1986} are slightly larger, in line with expected phase effects. For excitation, the experimental data available for particular channels \cite{Thorn2007JCP,Ralphs2013,Matsui2016} are scaled in order to estimate the total cross section \cite{deVera2021}, so as the data reproduces rather well the calculations. The cross section for the other major inelastic channel, namely dissociative electron attachment (DEA), is more difficult to calculate, and here is approximated from recommendations based on experimental data \cite{Itikawa2005}, shown in the figure by a dotted line.

As for the elastic scattering cross sections, they can be calculated \textit{ab initio} for water in the condensed phase. The Dirac-Hartree-Fock equation is solved by using a projected-potential approach based on Gaussian functions \cite{taioli2009surprises,Taioli2010,Morresi2018} for a cluster of six water molecules \cite{Taioli2021JPCL}, in an effort to approximate the liquid environment. The elastic cross sections are represented in Fig. \ref{fig:XSe}(b). Symbols correspond to experimental data for water vapour \cite{Itikawa2005,Katase1986,Cho2004a}, while the dotted line is that calculated for the water molecule within the Mott theory. The solid line depicts the Dirac-Hartree-Fock calculation for a cluster of six water molecules. While the calculation for the single water molecule reproduces fairly well the experiments, the results for the water cluster shows a clear deviation, which manifest the effect of the condensed phase on the elastic scattering.

\section{Track-structure simulations}

The cross sections described above are the information needed to simulate the secondary electron emission and transport by means of the Monte Carlo code SEED 
\cite{Taioli2021JPCL,Dapor2017PRB,DaporBook}, together with the electron-phonon and electron-polaron interactions, which are included by means of the Fr\"{o}hlich and Ganachaud-Mokrani models \cite{Taioli2021JPCL,Dapor2017PRB}.

\begin{figure}[t]
	\includegraphics[width=\columnwidth]{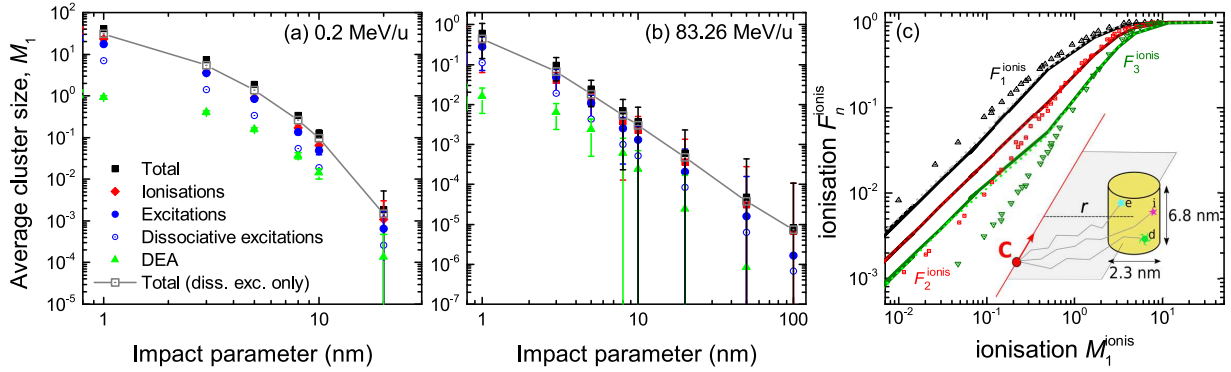}
	\caption[]{Clusters of inelastic events produced by carbon ions of different energies passing by some impact parameter from the centre of a cylinder of radius 2.3 nm and height 6.8 nm, similar to two DNA convolutions or 20 base pair. The inset in the right hand panel depicts the collision geometry. (a) Average size $M_1$ of the damage clusters produced by 0.2 MeV/u carbon ions, as a function of the impact parameter, by combinations of particular inelastic channels: ionisation, excitation and DEA. Open circles show the results for excitation clusters when only 40\% of them produce damage. Open squares connected by lines represent the total size of the damage cluster when only 40\% of excitations produce damage. (b) The same but for 83.26 MeV/u carbon ions. (c) Cumulative probabilities $F_n^{\rm ionis}$ to produce clusters of $n$ or more ionisations, as a function of the ionisation average cluster size $M_1^{\rm ionis}$. Here lines represent the results of simulations, while symbols correspond to experimental nanodosimetric data \cite{Conte2018}.}
	\label{fig:clusters}
\end{figure}

The program SEED is used to generate secondary electrons in liquid water along carbon ion paths of different energies, the latter passing by at different impact parameters from the centre of a sensitive nanocylinder, perpendicular to the path, with dimensions similar to two convolutions (or 20 base pairs) of a DNA molecule (2.3 nm diameter and 6.8 nm height). The collision geometry is depicted in the inset of Fig. \ref{fig:clusters}(c). Such a DNA-like target is commonly used in radiobiological studies.

\section{Results and discussion}

The Monte Carlo simulations allow calculating the average size of clusters of damaging events (ionisations, dissociative excitations, DEA) produced by carbon ions in the DNA-like volumes placed at different distances from the ions’ paths. The impact parameters analysed range from 1 to 100 nm, and the carbon ion energies have been varied between 0.2 MeV/u 
and 83.26 MeV/u. 
While initial energies of hundreds of MeV/u are typical in clinical carbon ion beams (so they present ranges of the order of tens of centimetres in tissue), the ions progressively lose their energy while traversing the body, going down to energies of around hundreds of keV/u around the Bragg peak region. Thus, 0.2 MeV/u corresponds to a typical energy of carbon ions around the Bragg peak (where their biological effects are most severe), while 83.26 MeV/u corresponds to a situation closer to the plateau region of the depth-dose curve. Such energetic carbon ions can also be found in cosmic radiation.

Figures \ref{fig:clusters}(a) and (b) represent the average size of damage clusters, $M_1$, as a function of the impact parameter, for these two limiting energies of 0.2 and 83.26 MeV/u. The clusters due to different physical mechanisms (ionisations, excitations and DEA) are represented by different symbols. It is known that only around 40\% of the electronic excitations are able to produce water molecule dissociation \cite{Thorn2008PhD}, so we only considered this proportion of ``dissociative excitations'' as damaging events. Open circles depict the average size of clusters produced by these dissociative excitations, while open squares connected by lines are the average sizes of the clusters due to the combination of ionisations, DEA and dissociative excitations. Error bars represent the statistical uncertainties obtained from a large number of simulations \cite{Taioli2021JPCL}. For 0.2 MeV/u, the average cluster size drops to zero for impact parameters larger than 20 nm, as secondary electrons are not able to reach large distances. For 83.26 MeV/u, the clusters are non negligible for longer distances, due to the production of more energetic delta electrons. Still, the average cluster sizes are much larger for 0.2 MeV/u than for 83.26 MeV/u, remarking the much larger relative biological effectiveness of carbon ions around the Bragg peak region. Figures \ref{fig:clusters}(a) and (b) also remarks the fact that ionisations are the main events contributing to the size of the damage clusters, followed by dissociative excitations, and, by far, by DEA. In fact, ionisations make up for around 75\% of the size of the clusters, while dissociative excitations contribute around 20\% and DEA only around 5\% \cite{Taioli2021JPCL}.  

Experimental nanodosimetry allows measuring the clustering of ionisation events in macroscopic gas-phase detectors having nanometric equivalent volumes of liquid water. Figure \ref{fig:clusters}(c) shows by symbols the distributions of the cumulative probabilities $F_n^{\rm ionis}$ to produce clusters of size equal or larger than $n$ (with $n = 1$, 2 or 3) as a function of the average ionisation cluster size $M_1^{\rm ionis}$ coming from a compilation of experimental data \cite{Conte2018}. It is worth to mention that such a curve is universal, and independent of the dimensions and particularities of the nanodosimeters. Calculated results coming from different ion energies and impact parameters, depicted by lines, reproduce experiments in a wide range of conditions, what validates the performance of the simulation code.




\section{Summary and conclusions}


As the direct damage of biological materials is determined by the clustering of inelastic events in volumes similar to the dimensions of a few convolutions of DNA, it is important to develop simulation tools capable to predict such clustered damage under irradiation. Given the current interest in applying carbon ion beams for cancer radiotherapy due to their increased relative biological effectiveness, and considering that liquid water is the main constituent of living tissue, we developed Monte Carlo simulations to study the clustered direct damage produced by carbon ion beams, in a wide range of energies relevant for hadrontherapy, in nanovolumes of liquid water.  Particular attention was paid to provide accurate \textit{ab initio} descriptions of the electronic excitation spectrum of liquid water as well as the elastic collisions, in order to feed the Monte Carlo code with reliable cross sections, so relevant to describe the propagation of the very low energy secondary electrons.

Time-dependent density functional theory (TDDFT) calculations yielded the ELF of liquid water in excellent agreement with the available experimental data. The ELF was extended to large energy and large momentum transfers by means of the MELF-GOS method and then used to obtain electronic cross sections within the dielectric formalism. The energy and angular distributions of secondary electrons produced by carbon ions reproduced well the known experiments. The same finding applies to the electronic excitation and ionisation cross sections for electrons when appropriate low energy corrections are implemented. Elastic scattering cross sections for liquid water were estimated by means of calculations for clusters of six water molecules by solving the Dirac-Hartree-Fock equation.

Simulations with the SEED code allowed to determine the clustering of inelastic events in cylinders of volume similar to two DNA convolutions, showing how for low carbon ion energies the average size of clusters can be hundreds of times larger than those for high energies far away from the Bragg peak region. Such clusters are mainly composed, independently of the carbon ion energy, by ionising events (up to 75--80\%), while DEA has a minor contribution. Moreover, the simulated cumulative distributions of ionisation clusters reproduce rather well the available measurements, ionisations being the inelastic events more easily accessible to experimental nanodosimetry.

\ack This project received funding from the European Union’s Horizon 2020 Research and Innovation programme under the Marie Sklodowska-Curie grant agreement no. 840752. M.D. and S.T. acknowledge the Bruno Kessler Foundation and the National Institute of Nuclear Physics for unlimited access to their computing facilities, and the Caritro Foundation for the grant High-Z ceramic oxide nanosystems for mediated proton cancer therapy. This work was also supported by the Spanish Ministerio de Ciencia e Innovación and the European Regional Development Fund (Project PGC2018-096788-B-I00); the Fundación Séneca-Agencia de Ciencia y Tecnología de la Región de Murcia (Project 19907/GERM/15); and the Ministry of Science and Higher Education of the Russian Federation as part of World-class Research Center program: Advanced Digital Technologies
(contract No. 075-15-2020-934 dated 17.11.2020).

\section*{References}
\bibliographystyle{iopart-num}
\bibliography{library}

\end{document}